# Title: Digitally Programmable Higher-order Harmonic Generation Using Nonlinear MEMS Resonator

**Authors: Gayathri Pillai[1] and Sheng-Shian Li[1,2]***

**Affiliations:**

[1]Institute of NanoEngineering and MicroSystems, National Tsing Hua University, Taiwan.

[2]Department of Power Mechanical Engineering, National Tsing Hua University, Taiwan.

*Correspondence to: ssli@mx.nthu.edu.tw

## Abstract

A fully electrically interfaced chip-scale Micro-Electro-Mechanical System based higher-order harmonic generator has multitudes of applications as it can simultaneously facilitate monolithic integration with on-chip electronics and offer microscopic footprint in comparison with the bulky and expensive high-order harmonic generation setup in optics. Here we report the first demonstration of a record-high harmonics generation over 100 tones using a thin-film Lead Zirconate Titanate flexural mode microresonator operating in an uncontrolled environment at CMOS compatible voltage levels. The high count of harmonics is achieved through its combined electromechanical and material nonlinearity. The resonator's harmonics can be befittingly controlled post-fabrication as well using multiple techniques such as modulating dynamic stiffness, altering the direction of the frequency sweep, tuning the driving frequency tone, and applying bias-voltage. Our experimental findings steer towards an attractive combination of digitally controllable features that can custom fit different applications for Internet-of-Things.

## Introduction

Nonlinear optical phenomena have been captivating the interests of researchers for over half a century. Some of the prominent domains of study such as frequency comb generation[1], parametric oscillation[2], and higher-order harmonic generation[3-5] (HHG) have been actively investigated because of their strong significance in multiple sectors of science and engineering such as spectroscopy[6], telecommunications[7], quantum information processing[8], etc. In particular, HHG in gas-phase laid the foundation for attosecond science[9] and recently solid-phase HHG[10] with good power handling was reported in bulk crystal that generated up to 25$^{th}$ harmonic. Several works have been done in optics to enhance the harmonics output such as engineered nanostructures[11,12], utilizing two-dimensional materials[13,14], photon acceleration technique[15], etc. Analytical works based on *Ab initio* simulation predict a possibility of generation up to 30 harmonics in solids[16] and over 100 in a free-standing monolayer[17]. However, most of the aforementioned approaches for HHG generation in optics involve a complex and expensive experimental setup that poses a bottleneck for the portable and cost-effective System-on-Chip (SoC) implementations.

Established semiconductor processing techniques are used to fabricate miniaturized MEMS resonators[18] using batch process schemes. This flexibility makes MEMS an attractive option from an application perspective. Phenomena equivalent to nonlinear optics have been demonstrated on electrically driven microscale mechanical devices[19-21] extensively. Phononic frequency combs formation due to the interaction between the primary mode and driven phonon mode was demonstrated in piezoelectric MEMS[22] recently which undoubtedly paves way for MEMS-based applications such as miniaturized resonant sensors[23], phonon computing[24], etc. The nonlinear attribute of MEMS structures has also been widely investigated to enhance the efficiency of energy harvesters by suitably programming its region of operation[25] in the nonlinear regime. HHG was demonstrated in capacitive MEMS-based comb drive resonators[26] that used the internal resonance phenomenon to couple energy from the fundamental mode to higher order mode. Through the 1:3 coupling of energy between the vibrational modes, electromechanical oscillators with enhanced noise performance was achieved[27].

Nevertheless, the count of HHG in capacitive MEMS is low and also a controlled environment is required for its operation. Here, we report the generation of more than 100$^{th}$ harmonic tone using a single Lead Zirconate Titanate (PZT) MEMS resonator excited at its fundamental mode of interest at complementary metal-oxide-semiconductor (CMOS) compatible voltage levels while being operated in air at room temperature. The HHG phenomenon was verified using both electrical and mechanical output frequency spectra. In addition to device engineering, several methods to control the region or point of nonlinear operation post-fabrication are explored using techniques such as dynamic stiffness control of the MEMS resonators[28], drive frequency sweep orientation[29], input frequency tone offset from fundamental resonance and bias-voltage tuning feature[30] of PZT. We demonstrate a flexural mode Thin Film Piezo-on-Silicon (TPoS) MEMS resonator-based HHG setup operating at its fundamental frequency, ~320kHz, that can generate up to 108$^{th}$ harmonic (35MHz). Meticulously chosen resonator mode shape and drive scheme along with the exceptional nonlinear feature of PZT[31] were utilized to achieve an extremely high count of overtones.

## Results

**Design Concept and Proposed Application.** Flexural modes possess large structural displacement and have been widely considered as a category of deformation which features lower power handling ability[32] and this attribute permits the resonator to be easily driven into nonlinearity. In this work, a PZT-on-Silicon-based MEMS resonator is designed to operate in flapping mode along a controllable nonlinear regime for HHG demonstration. Fig. 1(A) shows the schematic of the proposed HHG scheme using a fully electrically interfaced MEMS resonator. The resonator is a rectangular shaped suspended structure with two anchors to support the device and enable electrical signal routing to the resonator's active area. The device's material stack comprises a 2μm thick Silicon (Si), followed by a 0.15μm Platinum (Pt) and on top of which a 1μm transducer layer of PZT is deposited with a subsequent 0.3μm Gold (Au) electrode patterned as interdigitated electrodes. The input AC signal delivered via the input electrode actuates the resonator and the electrical signal is thereby converted to mechanical displacement of the suspended structure through the converse piezoelectric effect of PZT thin film. The charge generated upon the motion of the plate resonator is detected by the output top electrode. To generate HHG, a single-tone frequency input ($f_d$) close to the flapping mode resonant frequency is given to the resonator's input electrode with an amplitude high enough to drive the resonator into nonlinearity. Upon driving the resonator in the desired nonlinear configuration, strong integral multiples of $f_d$ can be observed in the spectrum of output electrode as shown in Fig. 1(A). The engineered mode shape along with drive level dependent nonlinear feature of PZT resulting from the material's domain wall motion enhances the efficiency of HHG[33]. Several digitally programmable modulation schemes which can meticulously control the HHG count as shown in Fig. 1(B) are also explored in detail in the following sections. Hence, the strong and controllable HHG phenomenon without the requirement of any complex ensemble makes PZT microresonator a compelling candidate for several applications such as frequency synthesizer and acousto-optic modulator (AOM) as depicted in Fig. 1(C). Two different resonator samples based on same mode shape but different operation frequency are investigated in this work (Fig. 5(C)). The resonant frequency of the suspended structure is dependent on the lateral dimensions and also the thickness of the device layers. A detailed resonator design guideline is presented in supplementary Fig. S1.

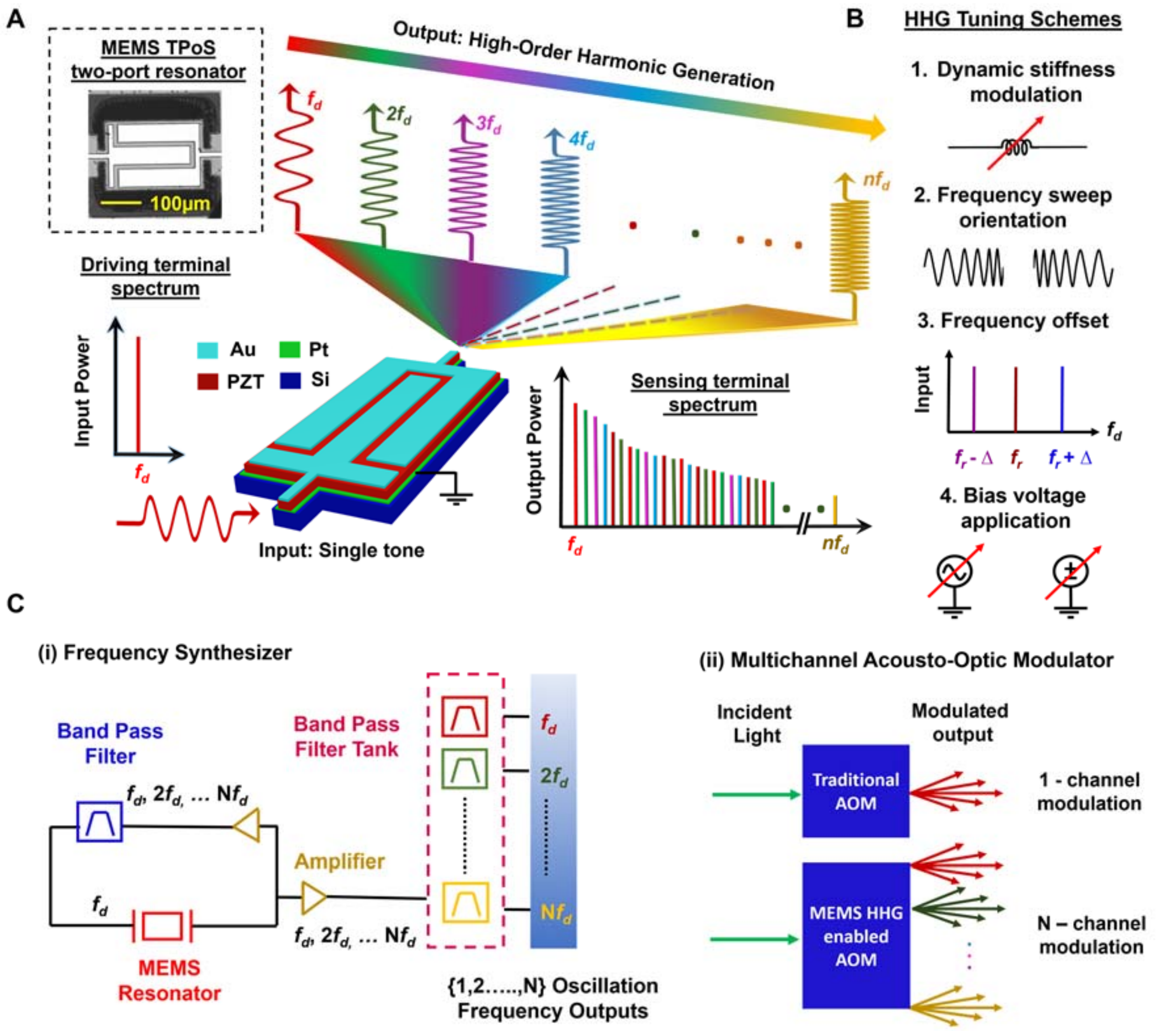

***Fig. 1. Overview of HHG scheme using single MEMS resonator.*** *(A) Schematic of a chip-scale HHG using a single frequency AC input with an amplitude sufficient to push the resonator into a nonlinear regime of operation. Inset shows the optical image of the resonator. (B) In addition to the ability of the MEMS system to generate harmonics, it has multiple overtone count control schemes. (C) Proposed applications including (i) Frequency synthesizer composed of a sole resonator that caters for multiple applications simultaneously rather than having an individual resonator for each application. (ii) Multichannel AOM utilizes the mechanical motion of the resonator to modulate the incident light source. Resonator motion is a function of resonator's operating point, thereby providing a range of modulation options.*

**Nonlinear MEMS Characterization.** The measured vibration pattern of the flapping mode and different drive/sense setups used to tune the nonlinear behavior of the resonator over a range of applied input voltages (and equivalent power) are shown in Fig. 2(A). At resonance, the average displacement is higher at the central portion when compared to the sides and hence when the center electrode is used to drive the resonator it is termed as low-stiffness drive, and high-stiffness drive when the side electrodes are used for actuation. A Zurich HF2LI Lock-in amplifier is used to drive and sense the resonator's electrical response (Fig. 6(B)). Due to a strong hardening feature, the resonator tends to go into a more nonlinear regime when the frequency is swept from low to high frequency. Driving a resonator in the low-stiffness region pushes the device into more nonlinearity than operating in the high-stiffness region as the effective stiffness for the former configuration is lesser than the later[34]. The resonant peak of Device A for the 36mV$_{pk}$ drive is consistent for all the drive configurations as it is in the linear region of operation. However, when the input is increased gradually, the nature of response for each of the configuration varies corresponding to varied nonlinear behaviors. Forward frequency sweep (low to high frequency) in a low-stiffness drive configuration shown in Fig. 2(B) has the highest output amplitude with maximum nonlinearity whereas a degenerate nonlinearity can be observed in Fig. 2(C) with the reverse frequency sweep (high to low frequency) in a high-stiffness drive configuration. Hence, it can be concluded from the sweep orientation study above that a low to high frequency sweep technique should be adopted to harness maximum nonlinearity for a device undergoing hardening and vice versa if the device were to exhibit strong softening behavior along the mode of interest. To confirm this Duffing nature, mechanical vibration response is monitored using a Laser Doppler Vibrometer (Fig. 6(C)) for the maximum nonlinear configuration as shown in Fig. 2(D) and the out-of-plane displacement were recorded. Doppler effect is utilized to attain the displacement information of the resonator surface. A power sweep for the forward frequency sweep was conducted and the displacement nature matches very closely with electrical readout trend of Fig. 2(B). This shows that the mechanical displacement of the microresonator is also modulated concurrently by the driving electrical signal and the device's stiffness configuration.

**Tunable HHG Measurement.** HHG phenomenon is initially explored using two input configurations: (i) a frequency sweep, and (ii) a single frequency tone. In the frequency sweep methodology, the input signal is swept along a narrow frequency range which includes the resonant frequency of the target mode shape. The direction of sweep is selected depending on the degree of nonlinearity requirement for any given application. On the other hand, for a single frequency tone input, the device is driven at a constant frequency that corresponds to a fixed degree of nonlinearity. Different points of operational frequency translate to unique points on the hysteresis loop and hence result in different harmonic or overtone generation.

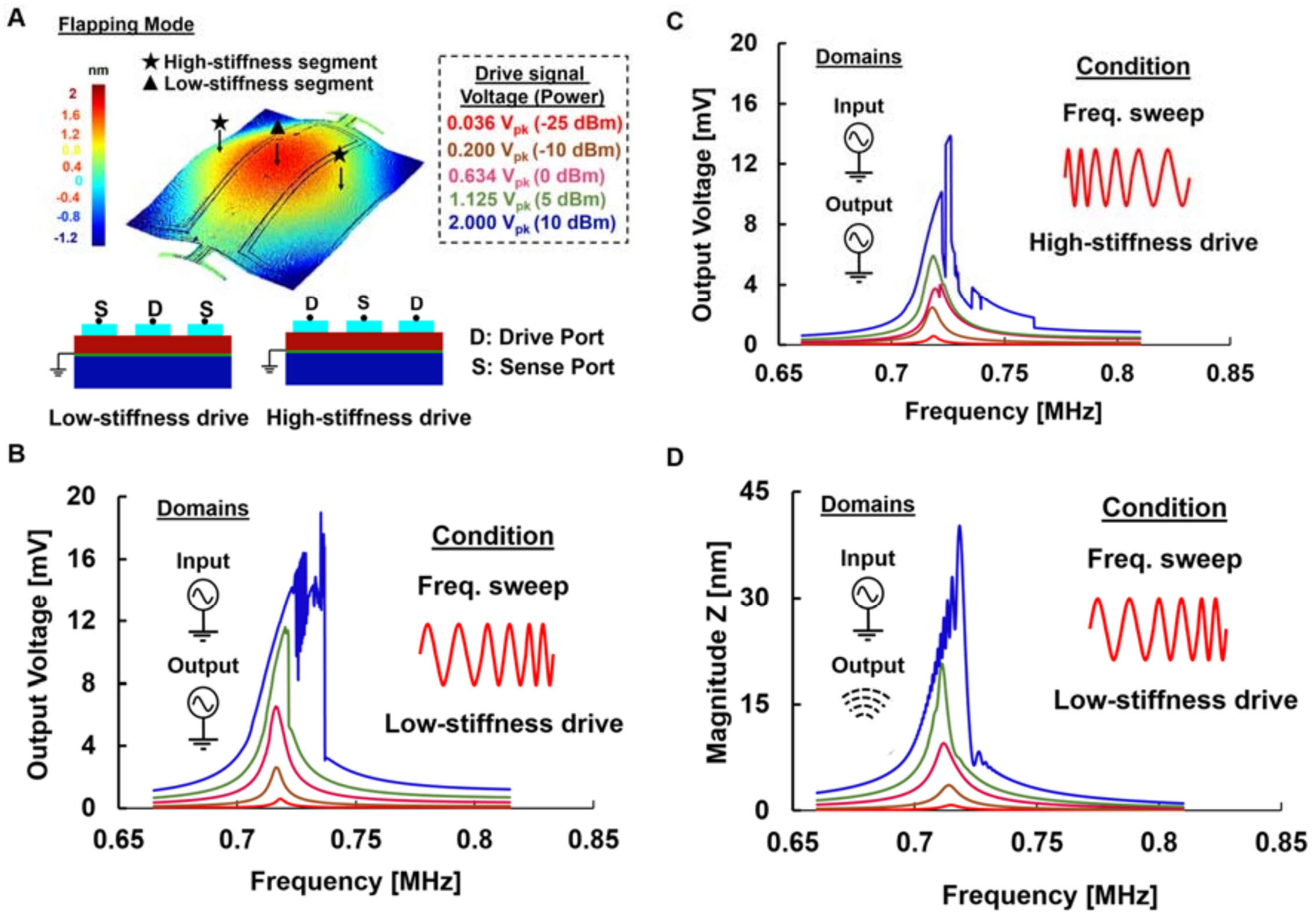

***Fig. 2. Electrical and mechanical frequency spectra of Device A.*** *(A) The mode shape of the resonator with different stiffness regions labeled. Drive/sense configurations employed for nonlinearity measurement of the resonator are also shown. The inset presents the voltage to power equivalency for all measurements in this work. (B) Forward frequency sweep for a low-stiffness drive. (C) Reverse frequency sweep for a high-stiffness drive. (D) Laser Doppler Vibrometer (LDV) out-of-plane motion measurement for forward frequency sweep in the low-stiffness configuration.*

For Fig. 3(A-C), the resonator is driven using a Lock-in amplifier while the output spectrum is measured in a Keysight Spectrum Analyzer (SA) (Fig. 6(B)). Fig. 3(A) shows the highly nonlinear drive/sense configuration operated in a forward frequency sweep of ±50kHz around the resonant frequency ($f_r$). For an input of -25dBm, only the fundamental mode can be observed which is indicative of the fact that there is no HHG phenomenon and that the resonator is still operating in the linear region. As the drive amplitude is increased, the HHG phenomenon starts unveiling, and at 0dBm the 2$^{nd}$ harmonic peak is observed. For a 10dBm drive, the HHG is very strong such that for an output frequency span of 0.5-10MHz all the 14 harmonics (odd and even) can be observed. To investigate the ability to control the harmonic generation by altering the drive/sense scheme, the resonator is driven in the high-stiffness configuration with a reverse frequency sweep (Fig. 3(B)). Since the mode undergoes hardening, reverse frequency sweep will yield a lower nonlinear frequency response. Corroborating the conclusions arrived in Fig. 2(B-C), the HHG counts were lesser in this condition as can be seen in Fig. 3(B). This measurement proves the strong digitally controllable harmonic feature of a MEMS-based HHG setup. Merely swapping the drive terminals and the orientation of the frequency sweep can suppress or generate harmonics to cater varying demands of multiplexing applications.

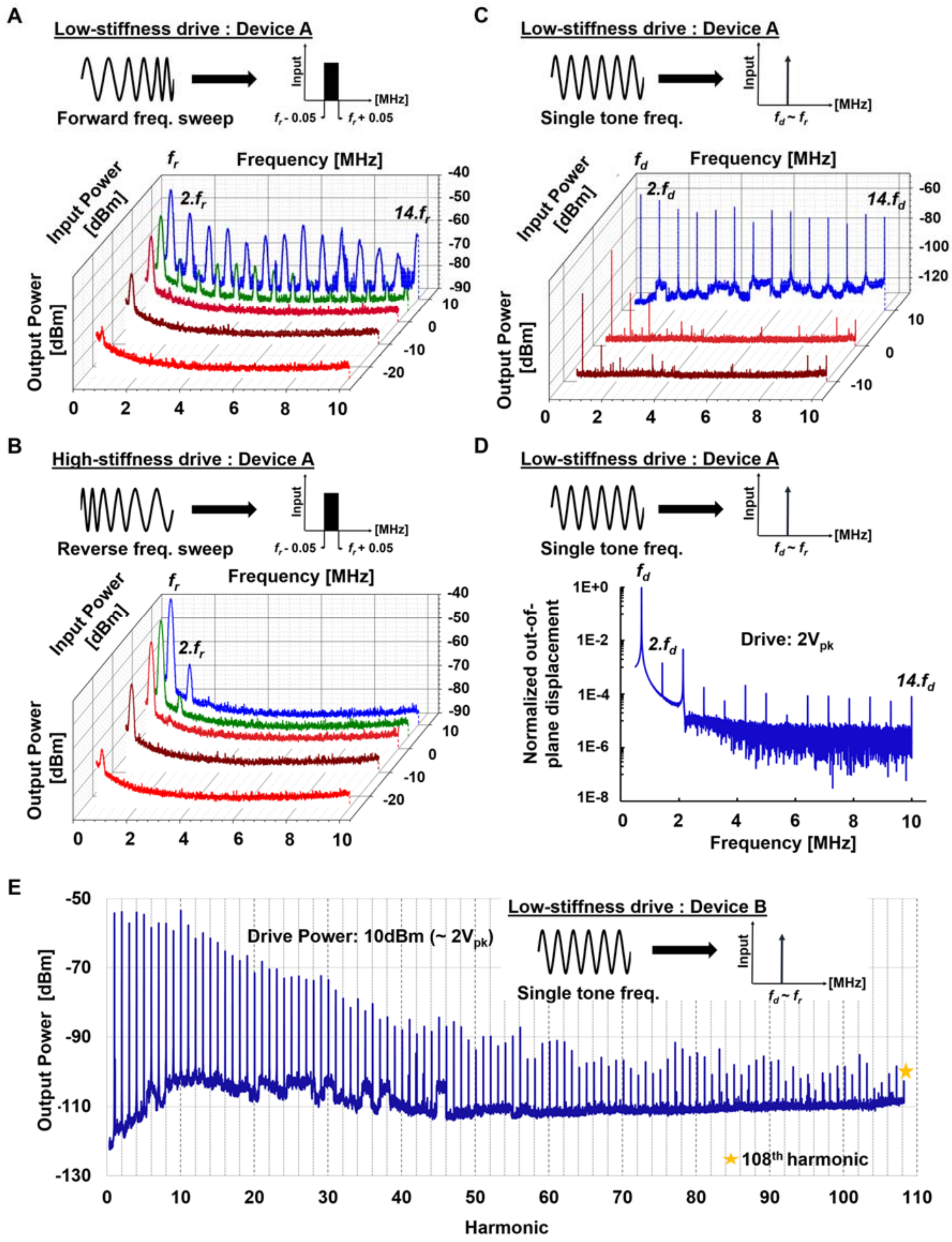

***Fig. 3. HHG using a single MEMS resonator.*** *Output spectra for Device A operating in the (A) high and (B) degenerate nonlinear regime for an input frequency sweep of* $f_r$ *±50kHz. The SA Bandwidth (BW) is set in Auto for the frequency sweep. (C) Output spectra of HHG for a single tone frequency input. The SA BW is set to 11Hz for a better Signal-to-Noise ratio. (D) The output spectrum of z-displacement for the highly nonlinear configuration for an electrical AC input at a frequency close to the flapping resonance mode. (E) The electrical output spectrum of HHG observation beyond 100 for Device B.*

Next on, the single-frequency tone driving condition is investigated and $f_d$ is chosen such that it is in the close vicinity of $f_r$ and the MEMS resonator is configured in a low-stiffness drive setup. Fig. 3(C) shows that the resonator can generate harmonics with just a single tone input and the number of overtones generated is the same as that of Fig. 3(A). This is due to the fact that for both the cases they are subjected to similar mechanical and electrical nonlinearity. To confirm that the harmonics are indeed mechanical modes, the output spectrum of LDV is investigated while being operated under the same drive condition as Fig. 3(C). Spectrum presented in Fig. 3(D) follows the same trend as the electrical output spectra, which confirms the structural displacement modulation and also HHG phenomenon when driven in a highly nonlinear regime using a single tone frequency.

To further extend HHG count, the MEMS geometry is altered so as to attain a lower dynamic stiffness for similar driving power. Device B is designed so as to have its flapping mode resonance at a lower frequency ($f_r$ ~320kHz). At a lower operational frequency, owing to a lower dynamic stiffness the resonator will exhibit a stronger nonlinearity than Device A. When being operated at $f_d$ around resonance for an input power of 10dBm configured in a highly nonlinear arrangement, a train of harmonics can be observed for a SA output frequency span of 0.1-35MHz as shown in Fig. 3(E) with the highest harmonic count being 108. *This work reports for the first time a record-breaking number of harmonic counts in an experimental demonstration of HHG!*

In addition to the above methods of controlling the HHG, we leverage the electronically enabled tuning ability of the harmonics. The idea for this approach is extracted from the fact that the count of the harmonic generation corresponds to the output amplitude of the frequency response. Besides the device configuration, the output's magnitude depends on the input drive frequency as well. When the device is driven into nonlinearity, it results in an interesting frequency behavior where the device has more than one output amplitude at a given frequency due to hysteresis as shown in the frequency response of Fig. 4(A) and (B). If the resonator is forced to operate in that particular bandwidth, the device voluntarily operates at a lower energy state as it is for any system when faced with multiple energy levels. As a result, in such cases, the device operates along the lower amplitude branch. This feature is exploited meticulously to attain a tailored number of HHG and for this, we use two techniques (i) alter the drive frequency tone and (ii) DC bias voltage tuning. Here these schemes of external tuning are demonstrated while the resonator is configured in a highly nonlinear regime for a constant input power of 10dBm. For the first tuning scheme, multiple drive frequency tones that map to different output amplitude were chosen as shown in Fig. 4(A). The output spectrum trace corresponding to the $f_d$ that has the maximum output voltage shows the strongest HHG phenomenon. At $f_r$+15kHz, the bidirectional frequency sweep of the suspended resonator yields two amplitudes for the same frequency. Hence, when driven at a constant frequency of $f_r$+15kHz, the device operates at point D and yields a lower HHG count than $f_r$-15kHz as the former maps to a lower amplitude than the later. DC bias voltage tuning of the resonant frequency[35] is next explored to change the point of operation in the nonlinear regime while maintaining the same $f_d$. This is based on the inherent polarization reversal mechanism of materials such as PZT[36]. The DC biasing has strong effects on the material's dielectric constant and piezoelectricity. This unique feature of the material provides yet another degree of freedom in this HHG study. Switching the polarity of the bias voltage generates an offset in the frequency response as illustrated in Fig. 4(B). Based on this observation, if the operation frequency and device configuration are predetermined, then the number of harmonics will eventually then be modulated by the bias dependent output amplitude. Hence building upon this concept, depending on the requirement of the harmonic counts at the output, suitable biasing can be applied to switch the operation point between the upper and lower branches as shown in Fig. 4(B). When the polarity of the DC bias was changed from -3V to +3V, the entire frequency response shifts resulting in $f_d$ to fall in the hysteresis region with dual amplitude (Points A and B). As explained earlier, the resonator will operate at point B as it is lower energy state and thereby yielding a lower harmonic count at the output. The mechanical mode shapes of the fundamental and the higher overtones that were recorded using the LDV scanning technique are shown in Fig. 4(C). The real time mode deformation for each of the modes when the fundamental mode is driven into significant nonlinearity are presented in the supplementary videos (Movies S1-3).

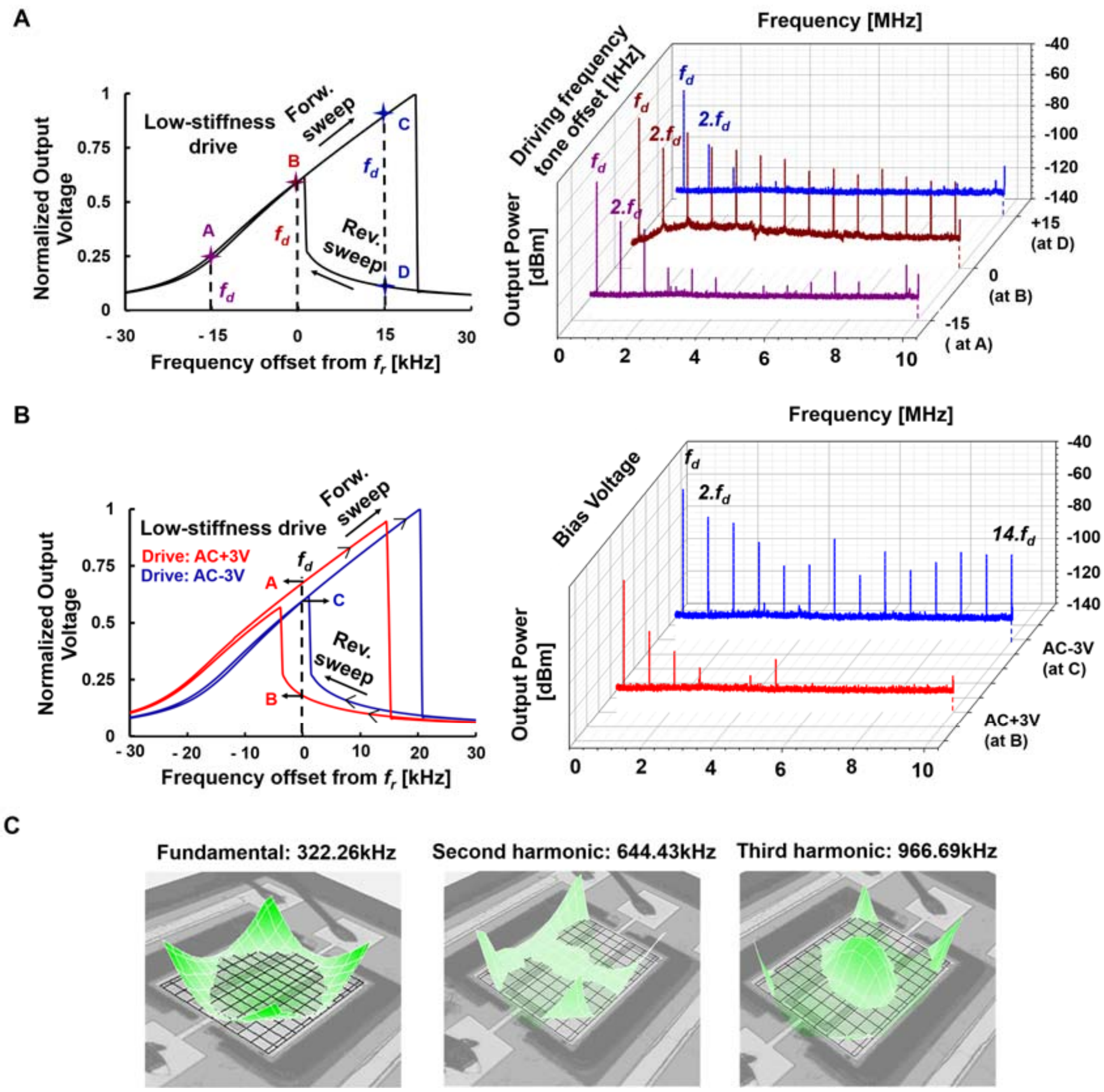

***Fig. 4. HHG control.*** *Output spectra of controlled HHG of Device A (A) corresponding to three frequencies having distinct output voltages. The bidirectional frequency sweep of the resonator along with the SA output spectrum is presented. (B) The frequency shift is DC voltage polarity dependent, hence for different DC bias voltages, different counts of harmonics can be achieved for the same* $f_d$*. (C) Measured mode shapes of fundamental, second and third harmonics of Device B generated using the out-of-plane displacement information acquired from LDV sensor head.*

## Discussion

The HHG experimental results illustrated in this work using PZT thin film demonstrates devices working in the sub-MHz range that successfully overcomes the hurdles in the conventional MEMS-based HHG techniques[26] such as high vacuum requirement, non-IC compatible DC voltages, weak electromechanical coupling, etc. Using the piezoelectric thin film based resonators, operation at other frequencies within the range where material nonlinearity is significant is also possible. We have observed a strong HHG in different resonator designs fabricated in the same batch operating at different fundamental modes even in the MHz range. To further extend this study, it is feasible to alter the resonator's material configuration and also the nonlinearity window of operation can be tailored by choosing a different primary mode of operation as the resonator's dynamic behavior varies with target resonant mode. Our work using PZT with its excellent electromechanical coupling coefficient supersedes not only in the quality and control of the harmonics but also attains a good signal to noise ratio for its harmonics as can be seen in Figures 3 and 4.

Albeit a highly efficient HHG system has been explored in detail here, there are many options still unexplored that can improve the nonlinear physics based system performance. A nonlinear frequency response similar to Fig. 2(B) has been observed in nanoelectromechanical resonators[37]; however, an HHG aspect wasn't explored by the authors. Hence, we strongly believe that HHG study can be extended to nanoscale resonators as well which would open up not only new engineering aspect but also unveil new dynamics in mechanical, electrical and material domains. Well established piezoelectric materials with strong coupling feature such as Aluminum Nitride, Lithium Niobate, etc. can be explored for extending the range of operation to even higher frequency domain. Aluminum Nitride based HHG system will enable monolithic integration of CMOS and MEMS[38] on a single chip which can cater to ever-growing demands of the Internet of Things.

In conclusion, through this work, we have successfully demonstrated the idea of a digitally controllable MEMS-based HHG system that can provide up to second-order overtones with a mere single AC signal when the resonator is operated in room temperature under atmospheric pressure condition. Several modulation techniques that can precisely control the nature of harmonics have been verified using both electrical and mechanical displacement spectrum. The extremely low power consumption (<10mW) of the system to generate harmonics is yet another notable feature of this PZT based HHG system. Crossing over domains to optics, nonlinear MEMS resonator with tunable HHG can also be a promising component to realize a multichannel AOM in optics as illustrated in Fig. 1(C). Currently, a single-channel modulator is used to modulate the incoming photon by utilizing the motion on the resonator surface at a certain operational frequency of the modulator component[39]. As the acoustic wave motion on the resonator surface generated by harmonic modes can also modulate the incident light, operating the modulator in the nonlinear regime of the resonator can efficiently provide an N-channel modulation in a chip-scale. Hence, this work that demonstrates high harmonic yield and post-fabrication tuning ability of MEMS resonator-based higher overtone generator sets a landmark in the field of HHG studies across multiple domains which have been actively investigated by researchers and engineers with an aim to translate the nonlinearity of material and/or structure to applications such as frequency synthesizers, sensors, modulation modules, frontend communication system, spectrometer, multiplexing systems, etc.

## Methods

### Microresonator Fabrication

The resonators were fabricated using a four-mask PZT Silicon-On-Insulator (SOI) microfabrication process provided by GlobalMEMS Co. Ltd as shown in Fig. 5(A). The process starts with a bare SOI wafer (Active-Si: 2µm/ Buried Oxide: 0.5µm/ Handle-Si: 400µm). Over the 2µm thick silicon, 150nm platinum is deposited. Platinum serves as the bottom electrode for the resonators. A 1µm PZT thin film is then deposited by the standard sol-gel process and then patterned to facilitate a through via to enable top and bottom electrode contact. A 300nm thick gold is sputtered over PZT and patterned using a lift-off process. Gold electrodes function as the input signal supply and output signal sensing medium for the devices and they are also routed suitably for probing pad interconnections purposes. The device geometry is defined using a dry etching process known as Reactive Ion Etching (RIE). Next on, backside Deep RIE is used to etch the 400µm thick Handle-Si. The Buried Oxide (BOX) serves as the etch stop layer for both the front and backside etching. Finally, dry etching of BOX thin film is performed to release the device. The Scanning Electron Microscope (SEM) image of device cross-section when it is cut along the thickness of the resonator is also presented in Fig. 5(A). The laser microscope image that maps the device profile under static condition is shown in Fig. 5(B).

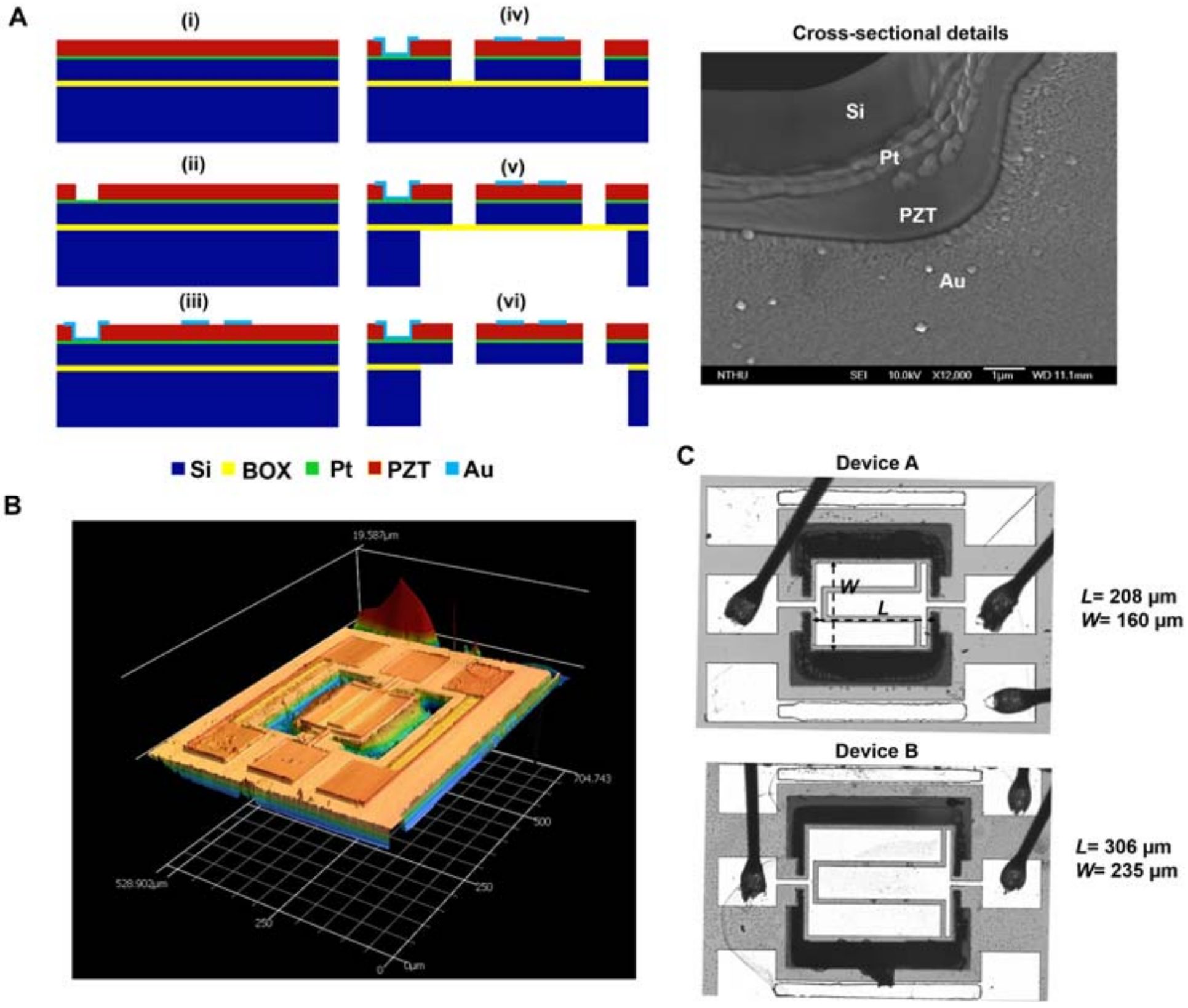

***Fig. 5. Device fabrication and analysis:*** *(A) Fabrication flow of the PZT TPoS MEMS resonator. (i) Incoming PZT thin film on SOI wafer. (ii) First mask: Pattern PZT. (iii) Second mask: Pattern top electrode. (iv) Third mask: Pattern PZT/Pt/SCS for resonator geometry definition. (v) Fourth mask: Backside etch for device release. (vi) Dry etch to remove the BOX. SEM image of the cross-sectional details of the device with all the layers labeled. (B) Profile of Design A measured using a Keyence laser microscope under static condition. (C) Optical microscope images of Devices A and B with their dimensions labeled.*

**Individual Device Details**

Different structural geometry and drive/sense signal configurations were explored to investigate the HHG phenomenon in MEMS TPoS resonators. The devices' optical images are presented in Fig. 5(C) and their individual details are stated below:

*Device A:$f_r$~715kHz*

The fundamental flapping mode resonant frequency for Design A is roughly around 715kHz and the rectangular resonator has released device dimensions of 208µm x 160µm. The device's resonant frequency may vary slightly from sample to sample because of the fabrication variation across the wafer. Asymmetric electrode design is used for the desired flapping mode i.e., the electrode coverages of drive/sense region over the active resonator structure are not equal. Ground-Signal-Ground (GSG) pads are designed to characterize the resonator.

*Device B: $f_r$ ~320kHz*

A larger version of Design A was designed to extend the nonlinearity hysteresis loop. The main resonant body dimensions are 306µm x 235µm. The drive/sense and ground electrodes are also scaled up accordingly to ensure enhanced transduction. Albeit the same $L/W$ ratio as Design A was maintained, the effective stiffness is expected to be lesser for Device B. The fundamental flapping mode occurs at a much lower drive frequency than Device A.

**Measurement Setup**

*Electrical Input - Electrical Output*

A complete electrical interface measurement setup was used to investigate the resonant behavior of the TPoS MEMS device. Fig. 6(A) shows the scheme where the device is operated using Agilent Network Analyzer E5071C. Network Analyzer (NA) is deployed to carry out initial resonator characterization and to extract the parameters relevant to the mode of interest. AC power is supplied to the device along with the desired frequency range and transmission plot is recorded in the NA. Next on, Zurich Instruments' HF2LI Lock-in was used as the signal source to perform bidirectional frequency sweep and also for the harmonic generation studies. The schematic for the aforementioned case is shown in Fig. 6(B). Keithley Sourcemeter was used to tune the DC bias potential of the resonator. To observe the harmonics in the frequency domain, the output of the resonator is connected to the Spectrum Analyzer.

*Electrical Input - Mechanical Output*

Laser Doppler Vibrometer (LDV) based non-contact vibration measurements were carried out to study the structural dependent nonlinearity and HHG features. An electrical AC driving signal was provided to the resonator's top electrode using the system's internal function generator. The Polytec MSA-100-3D LDV arrangement is based on the concept of Doppler effect i.e., sensing the frequency shift of backscattered light from a moving surface. Displacement magnitude and phase information of the fundamental mode and also higher harmonic modes for different driving power level, varied frequency sweep patterns, etc. were garnered using the setup shown in Fig. 6(C).

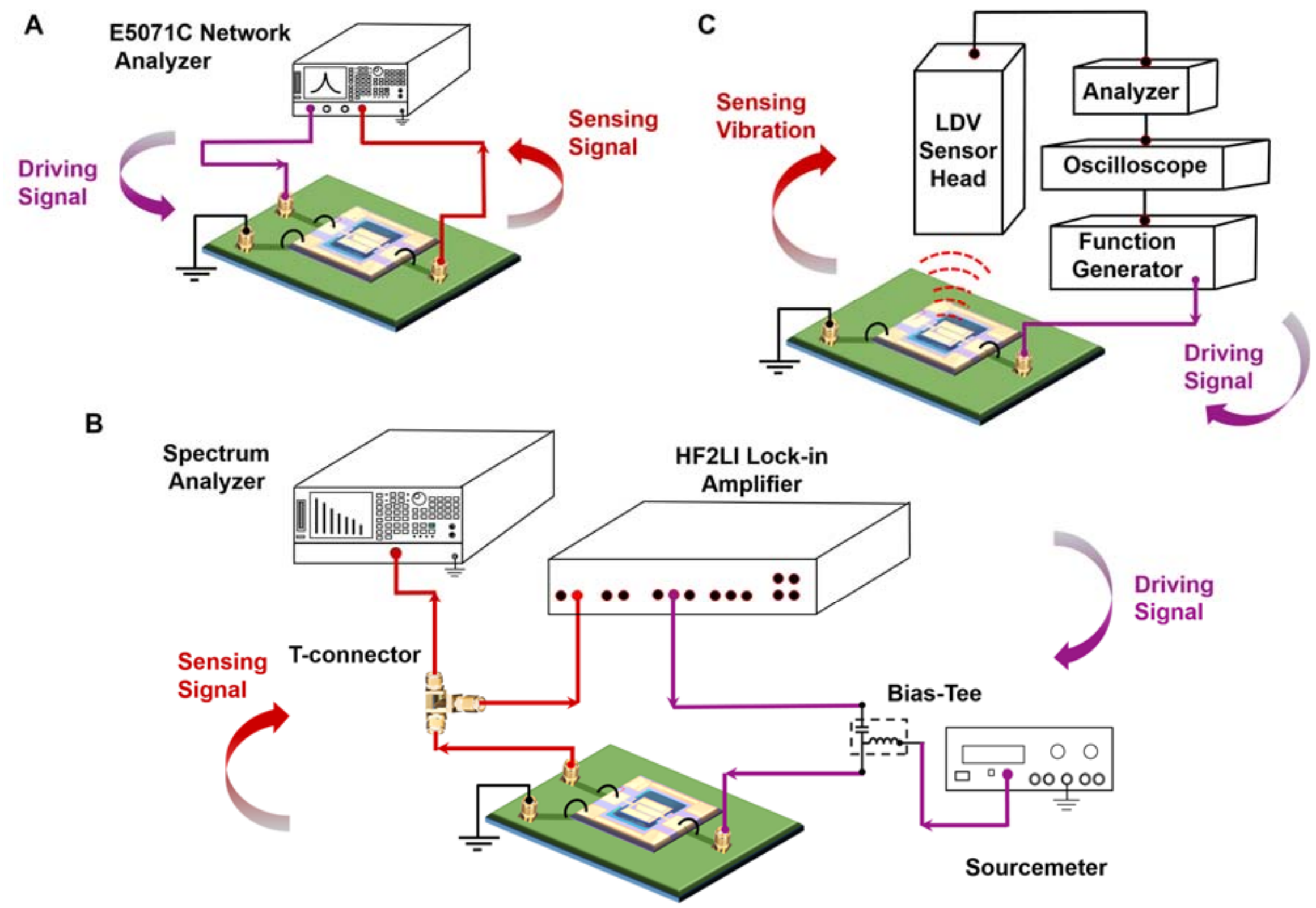

***Fig. 6. Measurement setup:*** *Electrical-in Electrical-out arrangement: (A) Agilent Network Analyzer serves as the time varying signal source. The driving power and the sweep frequency range can be precisely controlled using the Network Analyzer and it has a 50Ω termination impedance in each of its ports. (B) Zurich Instruments' HF2LI Lock-in amplifier's output is connected to the input of the resonator and the resonator's output is connected back to the Lock-in amplifier and also to the Spectrum Analyzer (SA) using a T-connector. SA is used to record the harmonic generation for different drive condition of the device powered by the Lock-in. The role of Lock-in amplifier is similar to the Network Analyzer. (C) Electrical-in Mechanical-out arrangement: An internal signal source of the LDV drives the resonator into resonance and the out-of-plane displacement of the resonator is recorded by the sensor head. Information in both time and frequency domain can be attained. In addition to the magnitude and phase details, graphical files of the mode shape vibration can also be generated which helps to analyze the fundamental and higher harmonic modes.*

**Acknowledgments:** The authors thank Prof. Gwo from National Tsing Hua University for the in-depth discussions on nonlinearity phenomenon and GlobalMEMS for the MEMS fabrication facility. This work was supported by the Ministry of Science and Technology (MOST), Taiwan under Grant MOST 107-2218-E-007 -020.

**Author contributions:** G. Pillai conducted the experiments and wrote the manuscript and S.-S. Li advised on all efforts.

**Competing interests:** Authors declare no competing interests. The work is under patent filing in both the US and Taiwan.

**Materials and Correspondence:** All the correspondences are to be made to Sheng-Shian Li.

**Data and materials availability:** All data are available in the main text or the supplementary materials.